\documentclass[fleqn,10pt]{wlscirep}

\title{Quantum Zeno effect in a nitrogen-vacancy center embedded in a spin bath}

\author[1]{Zhi-Sheng Yang}
\author[2]{Wen Yang}
\author[1]{Mei Zhang}
\author[1,*]{Qing Ai}
\author[1]{Fu-Guo Deng}

\affil[1]{Department of Physics, Applied Optics Beijing Area Major Laboratory,
Beijing Normal University, Beijing 100875, China}

\affil[2]{Beijing Computational Science Research Center, Beijing 100094, China}

\affil[*]{aiqing@bnu.edu.cn}


\begin{abstract}
We study the longitudinal relaxation of a nitrogen-vacancy (NV) center
surrounded by a $^{13}$C nuclear spin bath in diamond. By means of cluster-correlation expansion (CCE), 
we numerically demonstrate the decay process of electronic
state induced by cross relaxation at low temperature. It is shown
that the CCE method is not only capable of describing pure-dephasing
effect at large-detuning regime, but it can also simulate the quantum
dynamics of populations in the nearly resonant regime. We present
a proposal to slow down the decay of NV center via implementing quantum
Zeno effect (QZE). The numerical result shows that QZE can effectively
inhibit the decay of NV center.
\end{abstract}

\begin{document}

\flushbottom
\maketitle
%
%
\thispagestyle{empty}


Quantum Zeno effect (QZE) is the suppression of quantum evolution by
successive measurements \cite{Degasperis74,Misra77,Cook88,Itano90,Petrosky90}.
When a quantum system interacts with a bath, the quantum system undergoes
population/longitudinal relaxation and phase randomization (i.e., dephasing)
\cite{Nakazato96}, as determined by the overlap between the energy levels of
the system and the spectral density of the bath, i.e., the density of
states of the bath weighed by the system-bath couplings. Repetitive
measurement on the system effectively broadens its energy levels, which
can be understood from the Heisenberg energy-time uncertainty relation \cite%
{Kofman00}. This in turn changes the overlap between the system's broadened
energy levels and the bath spectral density. Consequently, the system's
decoherence could be either slowed down (i.e., the QZE) or accelerated
(i.e., the quantum anti-Zeno effect) \cite{Kofman00,Facchi01,Ai13}.

In 1990, QZE was experimentally observed in a two-level atomic system for
the first time \cite{Itano90}. By frequently interrupting the coherent
evolution of the atom with repetitive measurements, the atom was frozen in
its initial state. From then on, QZE has been successfully observed in
various physical systems \cite%
{Balzer00,Zhang03,Matsuzaki10,Xiao06,Bernu08,Xu11,Wolters13,Qiu15,Zhang11}.
Interestingly, the QZE has been applied to explain the quantum coherent
effects in biological systems, e.g., bird migration \cite{Kominis09} and
photosynthesis \cite{Kassal12}, in which the bath acts as a generalized
detector. The bath can monitor and even optimize the quantum evolution in
the relevant physiological processes \cite{Li16}.

In recent years, the nitrogen-vacancy (NV) center in diamond has attracted
broad interest from different disciplines \cite{Wrachtrup01,Wrachtrup06,Wang13,Song16-1}.
Because of its long coherence time at room temperature and easy
manipulations by optical pumping and microwave fields, the electronic spin
of the NV center becomes a promising candidate for quantum information
processing and quantum metrology \cite{Childress06,Dutt07,Zhao11-1,Zhou15,Zhou14,Song16-2}.
For ultrapure samples, the most
relevant bath is the $^{13}$C nuclear spins that randomly locate at the
lattice sites \cite{Zhao12}. These $^{13}$C nuclear spins are coupled to the
NV electronic spin through the hyperfine interaction and hence decoheres the
NV electron spin. The cross-relaxation between the NV electron spin and the $^{13}$C nuclear spins is usually strongly suppressed by the large energy
mismatch. In this regime, the NV electron spin decoherence is dominated by
pure dephasing (i.e., phase randomization) and various schemes have been
proposed and implemented to protect its quantum coherence \cite%
{Maze08,deLange10,Zhao11-2,Huang11,Zhao12,Yang15}. By contrast, there are
relatively fewer works on the cross-relaxation between the electron spin and
the nuclear spins \cite{Jacques09,WangH15,WangP15,WangP15b,Hall16}. In a
recent experiment \cite{Jarmola12}, it was discovered that the hyperfine
interaction with the nuclear spins dominates the electron spin relaxation at
sufficiently-low temperatures near the energy-level anticrossing of the NV
center. This experimental observation motivates us to suppress the NV
electron spin relaxation by the QZE. Previously, the quantum Zeno and
Zeno-like effects have been investigated both theoretically \cite{Qiu15} and
experimentally \cite{Wolters13} in an NV center without explicit
consideration of the $^{13}$C nuclear spin bath. However, the effect of the $^{13}$C bath on the QZE could be notable. A series of interesting phenomena
caused by the $^{13}$C nuclear spin bath have been predicted theoretically
and observed experimentally, e.g. anomalous decoherence effect \cite%
{Zhao11-2,Huang11} and atomic-scale magnetometry of distant magnetic
clusters \cite{Shi15,Zhao11-1}. Since the spin bath has a strong
non-Markovian nature, it can not generally be dealt with using the usual
quantum master equation approach in the Born-Markovian approximation. Among
the different methods for quantum dynamics of a central system in a spin
bath, the cluster-correlation expansion (CCE) \cite{Yang08,Yang09} approach
yields reliable pure-dephasing dynamics of the NV center and has been
successfully applied to predict anomalous decoherence effect in the NV
center, which was later observed experimentally \cite{Zhao11-2,Huang11}.
Therefore, the CCE approach is a good candidate for describing the QZE in an
NV center induced by the many-spin bath.

When a static magnetic field is applied to make the NV center close to the
level anticrossing point, the hyperfine interaction of the NV center with 
the $^{13}$C nuclear spins induces the NV electron spin relaxation. In this
paper, we generalize the CCE approach to describe the population dynamics of
an NV center embedded in the $^{13}$C nuclear spin bath. The numerical
simulation with the CCE approach yields convergent result and thus is capable
of fully describing the NV electron spin relaxation induced by the $^{13}$C
spin bath. We further demonstrate the QZE for the NV center surrounded by
the spin bath by optically reading out the NV center state at a frequency
faster than the NV spin relaxation. This optical manipulation provides an alternative route to combat the NV electron spin relaxation.

\section*{Results}

\subsection*{Model}

\begin{figure}[tbp]
\includegraphics[width=17cm]{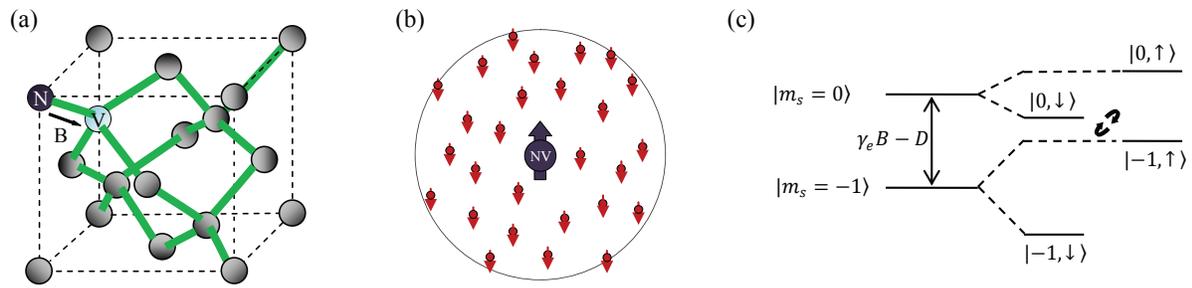}
\caption{Schematic illustration of QZE in an NV center surrounded by a spin
bath. (a) An NV center in the crystal structure of a diamond, where a static
magnetic field is applied along the principal axis of NV center. (b) The
bath of $^{13}$C nuclear spins and the NV center forms a quantum open
system. (c) The hyperfine structure of an NV center and one $^{13}$C nuclear
spin in the bath. }
\label{fig:Scheme}
\end{figure}

As shown in Fig.~\ref{fig:Scheme}, a negatively-charged NV center is coupled
to the $^{13}$C nuclear spins randomly located at the diamond lattices, with
the natural abundance of $^{13}$C being $1.1\%$ \cite{Zhao12}. Here, since
we focus on the longitudinal relaxation induced by $^{13}$C nuclear spins
far away from the NV center, in our numerical simulation we use the
randomly-generated bath in which there is no $^{13}$C nuclear spin in the
vicinity of NV center \cite{Zhao12}. The NV center consists of a
substitutional nitrogen atom adjacent to a vacancy. The ground state of NV
center is a triplet state with $m_{s}=0$ and $m_{s}=\pm1$ denoted by $%
\vert0\rangle$ and $\vert\pm1\rangle$ respectively. When there is no
magnetic field applied to the NV center, $\vert\pm1\rangle$ are degenerate
and they are separated from $\vert0\rangle$ by zero-field splitting $D=2.87$%
GHz \cite{Wrachtrup01,Wrachtrup06}. In an external magnetic field $B_z$
along the NV axis, the degeneracy between $m_{s}=\pm1$ levels are lifted and
thus the full Hamiltonian of the total system describing the NV center and
spin bath is
\begin{eqnarray}
H & = & H_{\mathrm{NV}}+H_{\mathrm{bath}}+H_{\mathrm{int}},  \label{eq:H}
\end{eqnarray}
where the Hamiltonian of the NV center has the form
\begin{eqnarray}
H_{\mathrm{NV}} & = & DS_{z}^{2}-\gamma_{\mathrm{e}}B_{z}S_{z}
\end{eqnarray}
with $\gamma_{\mathrm{e}}=-1.76\times10^{11}$ rad$\cdot$s$^{-1}$T$^{-1}$
being the electronic gyromagnetic ratio \cite{Zhao12} and $%
S_{z}=\vert1\rangle\langle1\vert-\vert-1\rangle\langle-1\vert$.

The Hamiltonian of the bath reads
\begin{equation}
H_{\mathrm{bath}}=-\gamma_{\mathrm{c}}B_{z}\sum_{i}I_{i}^{z}+\sum_{i<j}D_{ij}%
\left[\mathbf{\mathbf{\mathbf{I}}}_{i}\cdot\mathbf{\mathbf{I}}_{j}-\frac{%
3\left(\mathbf{\mathbf{I}}_{i}\cdot\mathbf{r}_{ij}\right)\left(\mathbf{%
\mathbf{r}}_{ij}\cdot\mathbf{\mathbf{I}}_{j}\right)}{r_{ij}^{2}}\right],
\end{equation}
where the gyromagnetic ratio of $^{13}$C nuclear spin is \cite{Zhao12} $%
\gamma_{\mathrm{c}}=6.73\times10^{7}$rad$\cdot$s$^{-1}$T$^{-1}$,
\begin{equation}
D_{ij}=\frac{\mu_{0}\gamma_{\mathrm{c}}^{2}}{4\pi r_{ij}^{3}}\left(1-\frac{3%
\mathbf{\mathbf{r}}_{ij}\mathbf{\mathbf{r}}_{ij}}{r_{ij}^{2}}\right)
\end{equation}
is the magnetic dipole-dipole interaction between the $i$th and $j$th $^{13}$%
C nuclear spins with $\mathbf{\mathbf{r}}_{ij}$ being the displacement
vector from $i$th to $j$th spins and $\mu_{0}$ being the vacuum permeability.

The electron spin and nuclear spins are coupled by hyperfine interaction.
Because we are interested in a large number of $^{13}$C nuclear spins far
away from the NV center, the hyperfine interaction is mainly described by
magnetic dipole-dipole interaction and thus the interaction Hamiltonian is
\begin{equation}
H_{\mathrm{int}}=\mathbf{S}\cdot\sum_{i}A_{i}\cdot\mathbf{\mathbf{I}}_{i},
\end{equation}
where
\begin{equation}
A_{i}=\frac{\mu_{0}\gamma_{\mathrm{c}}\gamma_{\mathrm{e}}}{4\pi r_{i}^{3}}%
\left(1-\frac{3\mathbf{\mathbf{r}}_{i}\mathbf{\mathbf{r}}_{i}}{r_{i}^{2}}%
\right)
\end{equation}
is the hyperfine coupling tensor, $\mathbf{r}_i$ is the position vector of $%
i $th nuclear spin, and the position of NV center is chosen as the origin.
Notice that we do not use the rotating-wave approximation which might
influence the existence of quantum anti-Zeno effect \cite{Zheng08,Ai10}.

\subsection*{Decay of NV Center Induced by Nuclear Spin Bath}

In this subsection, we discuss the longitudinal relaxation of electron
spin of NV center induced by the nuclear spin bath. When the energy gap
between $\vert0\rangle$ and $\vert-1\rangle$ approaches the energy gap of $%
^{13}$C nuclear spins, e.g. by tuning the magnetic field, the electron spin
exchanges polarization with nuclear spins, cf. Fig.~\ref{fig:Scheme}(c), and
the longitudinal relaxation of electronic spin occurs. For simplicity, we
assume that the decay of electron spin is mainly due to hyperfine interaction
with nuclear spins, and the effect of diamond lattice phonons on the
relaxation is beyond the scope of the present investigation.

The initial state of electronic spin can be optically initialized into $%
|0\rangle $, corresponding to the density matrix
\begin{equation}
\rho _{\mathrm{NV}}=|0\rangle \langle 0|.
\end{equation}%
In principle, the $^{13}$C nuclear spin bath should be taken as in the
completely random thermal equilibrium state since the typical experimental
temperature is much higher than the nuclear spin Zeeman splitting, even in a
strong magnetic field. However, previous calculations show that the results
from a thermal state for the nuclear spin bath are usually the same as those
from a randomly chosen pure state for the bath \cite{Yang09}. Therefore,
without loss of generality, we take the density matrix of the $^{13}$C bath
as
\begin{equation}
\rho _{\mathrm{bath}}=\prod_{i=1}^{\otimes N}|\downarrow \rangle _{i}\langle
\downarrow |,
\end{equation}%
where $|\uparrow \rangle _{i}$ ($|\downarrow \rangle _{i}$) is the spin-up
(spin-down) state of the $i$th $^{13}$C nuclear spin, with the quantization
axis along the N-V symmetry axis. As a result, the density matrix of total
system at the initial time is
\begin{equation}
\rho \left( 0\right) =\rho _{\mathrm{NV}}\otimes \rho _{\mathrm{bath}}.
\end{equation}%
The evolution of the coupled system is given by
\begin{equation}
\rho \left( t\right) =e^{-iHt}\rho \left( 0\right) e^{iHt}.
\end{equation}%
By partially tracing over the degrees of freedom of the bath, we could
obtain the survival probability of the initial state $|0\rangle $ of the NV
electron spin as
\begin{equation}
P(t)=\mathrm{Tr}_{B}\langle 0|e^{-iHt}\rho \left( 0\right) e^{iHt}|0\rangle .
\label{eq:P}
\end{equation}

\begin{figure}[tbp]
\includegraphics[width=15cm]{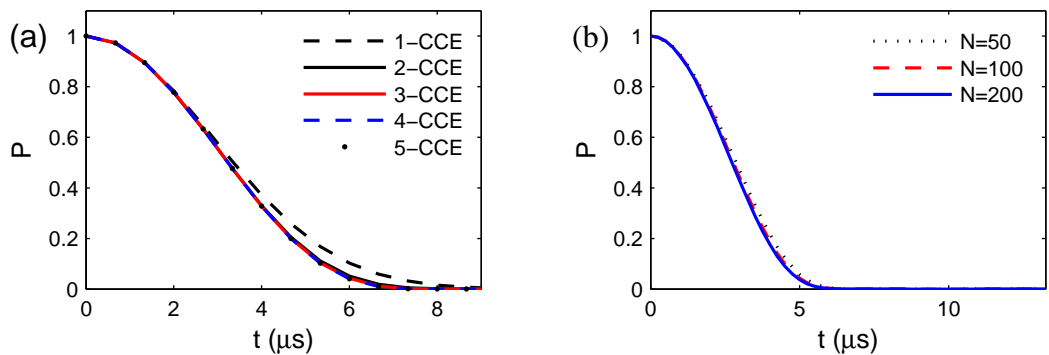}
\caption{The longitudinal relaxation process of the NV center. Survival
probability of initial state of electron spin by (a) different orders of CCE
with $N=100$; (b) $4$-CCE with $N=50$ (dotted line), and $N=100$ (dashed
line), and $N=200$ (solid line). In all cases, the magnetic field is set as $B=1024.98$Gs. }
\label{fig:Convergence}
\end{figure}

For a small spin bath, we can exactly calculate the longitudinal relaxation
of electron spin via Eq.~(\ref{eq:P}). However, with increasing size of the
nuclear spin bath, this approach quickly becomes unfeasible, because the
dimension of the Hilbert space grows exponentially with the number of nuclei
\cite{Dobrovitski09,Zhang07}. In this case, an approximate many-body theory
with high performance has to be considered. In previous works, the CCE
theory has been successfully applied to describe the pure dephasing of a
central spin in a spin bath \cite{Yang08,Yang09,Zhao11-1,Zhao11-2}. Here, we
generalize the CCE theory to deal with the longitudinal relaxation of the
electron spin in a spin bath. In Fig.~\ref{fig:Convergence}(a), we
numerically simulate the survival probability of the initial state of
electron spin surrounded by $N=100$ nuclear spins. When the order of CCE
method is increased, the results quickly converge, e.g., the results from
4th-order CCE and 5th-order CCE are indistinguishable. This suggests that
4th-order CCE already offers a reliable solution to the quantum dynamics of
central spin under the influence of a many-spin bath. Furthermore, we
explore the effect of the size of the bath on the decoherence. In Fig.~\ref%
{fig:Convergence}(b), the population dynamics of the electron spin is
investigated for $^{13}$C baths of different sizes. For a bath with $N=50$
nuclear spins, the survival probability already experiences an
exponential-like decay. When the size of the bath is doubled, the difference
before and after the change is negligible. And further enlarging the size
will not increase but decrease the difference. Because 4-CCE theory with a
bath of $N=100$ nuclear spins already yields a reliable result, it will be
adopted in the following investigations.

\subsection*{Quantum Zeno Effect}

\begin{figure}[tbp]
\includegraphics[width=17cm]{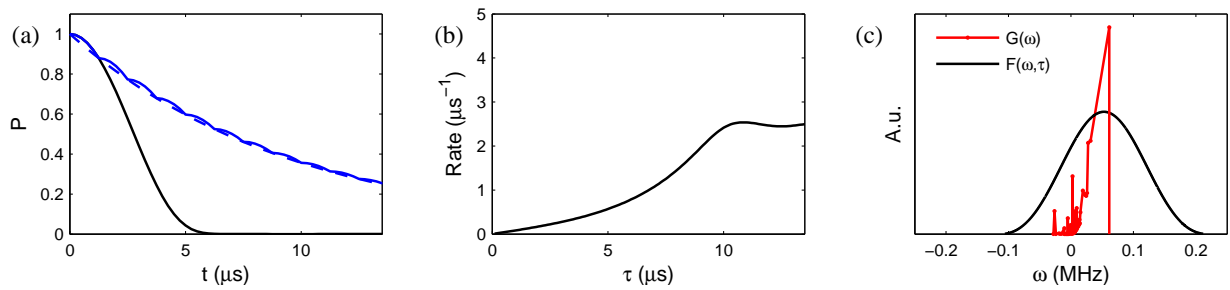}
\caption{The QZE in an NV center surrounded by a nuclear spin bath. (a) The
black full line denotes the undisturbed decay of NV center. The blue solid
line represents the survival probability under repetitive measurements,
while the blue dashed line is fitting it by exponential decay. (b) The
effective decay rate $R_{\text{eff}}(\protect\tau)$ vs the measurement
interval $\protect\tau$. (c) The spectral density $G(\protect\omega)$ and
measurement-induced level broadening $F(\protect\omega,\protect\tau)$ with $%
\protect\tau=12\protect\mu$s. }
\label{fig:QZE}
\end{figure}

In the QZE, the transitions between quantum states can be inhibited by
frequent measurements. After $N$ measurements, the survival probability of
the initial state is \cite{Facchi01}
\begin{equation}
P^{(N)}(t)=P\left(\tau\right)^{N}=e^{-R_{\text{eff}}(\tau)t},
\end{equation}
where $t=N\tau$ is the total duration and $\tau$ is time interval between
two successive measurements. The effective decay rate reads
\begin{equation}
R_{\text{eff}}(\tau)=-\frac{1}{\tau}\text{ln}P(\tau).
\end{equation}
In Fig.~\ref{fig:QZE}(a,b), we numerically simulate the population dynamics
by means of CCE method. In the short-time regime, the survival probability
reveals a quadratic decay as confirmed by the linear dependence of $R_{\text{eff}}(\tau)$ on $\tau$. Afterwards, as the survival probability quickly diminishes, the effective rate gradually approaches the steady value, which suggests an exponential decay for the population dynamics in the long-time limit. Therefore, if the free evolution of the electron spin under the influence of a many-spin bath is interrupted by repetitive measurements
before its behavior reaches steady state, the central spin will be
maintained in the initial state and the QZE occurs.

To understand the physical mechanism, we resort to the quantum Zeno and
anti-Zeno effects for an open system. In Ref.~\citen{Kofman00}, it was
illustrated that the effective decay rate of a central system in a bath of
harmonic oscillators is the overlap integral of the spectral density and
measurement-induced level broadening, i.e.
\begin{equation}
R_{\text{eff}}=2\pi \int_{0}^{\infty }F(\omega ,\tau )G(\omega )d\omega .
\label{eq:Reff}
\end{equation}%
The measurement-induced level broadening describes the broadening of energy
level of central system due to frequent measurements and is defined as
\begin{equation}
F(\omega ,\tau )=\frac{\tau }{2\pi }\text{sinc}^{2}\frac{(\omega -\omega
_{a})\tau }{2}
\end{equation}%
with $\omega _{a}=D+\gamma _{e}B_{z}$ being the level splitting between
electronic states $|0\rangle $ and $|-1\rangle $. For a bath of many spins,
because the spins are discrete, the spectral density is summarized over the
nuclear spins as
\begin{equation}
G(\omega )=\sum_{j}|\langle 0,\downarrow
_{j}|(A_{j}^{xx}+A_{j}^{yy})|-1,\uparrow _{j}\rangle |^{2}\delta (\omega
-\omega _{j})=\sum_{j}\frac{9\mu _{0}^{2}\gamma _{c}^{2}\gamma
_{e}^{2}(x_{j}^{2}+y_{j}^{2})}{256\pi ^{2}r_{j}^{10}}\delta (\omega -\omega
_{j}),
\end{equation}%
where the level splitting of $j$th nuclear spin
\begin{equation}
\omega _{j}=-\gamma _{c}B_{z}-\frac{\mu _{0}\gamma _{c}\gamma _{e}}{4\pi
r_{j}^{2}}(1-\frac{3z_{j}^{2}}{r_{j}^{2}})
\end{equation}%
has been modified due to the hyperfine coupling, ($x_{j},y_{j},z_{j}$) are
the three components of position vector $\vec{r}_{j}$ of $j$th nuclear spin.
Here we remark that because the nuclear spins are polarized at the initial
stage, the spin bath under consideration is equivalent to a bath of harmonic
oscillators. Clearly as illustrated in Fig.~\ref{fig:QZE}(c), the peak of $%
G(\omega )$ is very close to the peak of $F(\omega ,\tau )$. When the
interval $\tau $ between successive measurements on electron spin is
relatively long, i.e. in the long-time limit of Eq.~\ref{eq:Reff}, $F(\omega
,\tau )$ becomes a delta function, leading to maximal overlap between $%
F(\omega ,\tau )$ and $G(\omega )$ and hence maximum decay rate of the
electron spin $R_{\text{eff}}=2\pi G(\omega _{a})$. When the measurement
interval $\tau $ decreases, the delta function $F(\omega ,\tau )$ broadens,
so its overlap with the bath spectrum $G(\omega )$ and hence the electron
spin decay rate also decrease. By measuring the electron spin with a
sufficiently rapid frequency, the electron spin relaxation can be suppressed.

\section*{Discussion}

In this paper, we generalize the CCE approach to deal with the longitudinal
relaxation of a central electron spin due to its coupling to a 
nuclear spin bath in an NV center. The decay of electronic initial state is
Gaussian in the short-time regime, while it becomes exponential on longer
time scales. By interrupting the electron spin relaxation process with
successive measurements, the decay can be slowed down, with the effective
decay rate being determined by the overlap integral between the bath
spectral density and measurement-induced electron spin energy level
broadening \cite{Kofman00}. This gives rise to the QZE for the NV center
electron spin.

Previously, the CCE method has been applied to simulate the pure-dephasing
process of electron spin coupled to a many-spin bath in the far-detuned
regime. In this paper, the CCE method has been generalized to describe the
longitudinal relaxation process of the NV center in the near 
resonant regime.

Furthermore, we should also remark that in the previous investigations \cite%
{Wolters13,Qiu15} the QZE were demonstrated in an NV center as a close
system, while in this paper the QZE is demonstrated in an open quantum
system with a many-spin bath. Besides, the quantum Zeno-like effect in Ref.~%
\citen{Qiu15} is different from the QZE in that the quantum Zeno-like effect
originates from the Brouwer fixed-point theorem \cite{Layden15} and thus
occurs at particular measurement intervals.

\section*{Methods}

\textbf{The cluster-correlation expansion.} Here we generalize the CCE
method \cite{Yang08,Yang09} to treat electron spin relaxation in a spin
bath. Assuming that the bath consists of spin $i$ only, $P(t)$ is explicitly
calculated as
\begin{equation}
\tilde{P}_{\{i\}}=P_{\{i\}}\equiv \frac{\text{Tr}(\rho
(0)e^{iH_{\{i\}}t}|0\rangle \langle 0|e^{-iH_{\{i\}}t})}{\text{Tr}(\rho
(0)e^{iH_{\mathrm{NV}}t}|0\rangle \langle 0|e^{-iH_{\mathrm{NV}}t})},
\label{eq:1CCE}
\end{equation}%
where
\begin{equation}
H_{\{i\}}=H_{\mathrm{NV}}-\gamma _{\mathrm{c}}B_{z}I_{i}^{z}+\mathbf{S}\cdot
A_{i}\cdot \mathbf{\mathbf{I}}_{i}
\end{equation}%
is obtained from Eq.~(\ref{eq:H}) by dropping all terms other than spin $i$.

Assuming that the bath consists of spin $i$ and spin $j$, $P(t)$ reads
\begin{equation}
P_{\{i,j\}}\equiv\frac{\text{Tr}(\rho(0)
e^{iH_{\{i,j\}}t}\vert0\rangle\langle0\vert e^{-iH_{\{i,j\}}t})} {\text{Tr}%
(\rho(0) e^{iH_{\mathrm{NV}}t}\vert0\rangle\langle0\vert e^{-iH_{\mathrm{NV}%
}t})},  \label{eq:2CCE}
\end{equation}
where
\begin{equation}
H_{\{i,j\}}=H_{\mathrm{NV}}-\gamma_{\mathrm{c}}B_{z}\sum_{\alpha=i,j}I_{%
\alpha}^{z} +D_{ij}\left[\mathbf{\mathbf{\mathbf{I}}}_{i}\cdot\mathbf{%
\mathbf{I}}_{j}-\frac{3\left(\mathbf{\mathbf{I}}_{i}\cdot\mathbf{r}%
_{ij}\right)\left(\mathbf{\mathbf{r}}_{ij}\cdot\mathbf{\mathbf{I}}_{j}\right)%
}{r_{ij}^{2}}\right] +\mathbf{S}\cdot \sum_{\alpha=i,j}A_{\alpha}\cdot%
\mathbf{\mathbf{I}}_{\alpha}
\end{equation}
is calculated from Eq.~(\ref{eq:H}) by dropping all terms other than spin $i$
and spin $j$, and the spin-pair correlation is
\begin{equation}
\tilde{P}_{\{i,j\}}\equiv \frac{P_{\{i,j\}}}{\tilde{P}_{\{i\}}\tilde{P}%
_{\{j\}}}.
\end{equation}

Assuming that the bath consists of three spins $i$ and $j$ and $k$, $P(t)$
in this case becomes
\begin{equation}
P_{\{i,j,k\}}\equiv\frac{\text{Tr}(\rho(0)
e^{iH_{\{i,j,k\}}t}\vert0\rangle\langle0\vert e^{-iH_{\{i,j,k\}}t})} {\text{%
Tr}(\rho(0) e^{iH_{\mathrm{NV}}t}\vert0\rangle\langle0\vert e^{-iH_{\mathrm{%
NV}}t})},  \label{eq:3CCE}
\end{equation}
where
\begin{equation}
H_{\{i,j,k\}}=H_{\mathrm{NV}}-\gamma_{\mathrm{c}}B_{z}\sum_{\alpha=i,j,k}I_{%
\alpha}^{z} +\sum_{\alpha=i,j,k}\sum_{\beta>\alpha}D_{\alpha\beta}\left[%
\mathbf{\mathbf{\mathbf{I}}}_{\alpha}\cdot\mathbf{\mathbf{I}}_{\beta}-\frac{%
3\left(\mathbf{\mathbf{I}}_{\alpha}\cdot\mathbf{r}_{\alpha\beta}\right)\left(%
\mathbf{\mathbf{r}}_{\alpha\beta}\cdot\mathbf{\mathbf{I}}_{\beta}\right)}{%
r_{\alpha\beta}^{2}}\right] +\mathbf{S}\cdot
\sum_{\alpha=i,j,k}A_{\alpha}\cdot\mathbf{\mathbf{I}}_{\alpha}
\end{equation}
is calculated from Eq.~(\ref{eq:H}) by dropping all terms other than the
three spins $i$ and $j$ and $k$, and the three-spin correlation is
\begin{equation}
\tilde{P}_{\{i,j,k\}}\equiv \frac{P_{\{i,j,k\}}}{\tilde{P}_{\{i\}}\tilde{P}%
_{\{j\}}\tilde{P}_{\{k\}}\tilde{P}_{\{i,j\}}\tilde{P}_{\{j,k\}}\tilde{P}%
_{\{k,i\}}}.
\end{equation}

For the bath with arbitrary number of spins, $P(t)$ is generalized as
\begin{equation}
P_{\mathfrak{c}}\equiv\frac{\text{Tr}(\rho(0) e^{iH_{\mathfrak{c}%
}t}\vert0\rangle\langle0\vert e^{-iH_{\mathfrak{c}}t})} {\text{Tr}(\rho(0)
e^{iH_{\mathrm{NV}}t}\vert0\rangle\langle0\vert e^{-iH_{\mathrm{NV}}t})},
\label{eq:MCCE}
\end{equation}
where $H_{\mathfrak{c}}$ is obtained from Eq.~(\ref{eq:H}) by dropping all
terms other than spins in the cluster $\mathfrak{c}$, and the spin-cluster
correlation is
\begin{equation}
\tilde{P}_{\mathfrak{c}}\equiv \frac{P_{\mathfrak{c}}}{\prod_{\mathfrak{c}%
^\prime\subset\mathfrak{c}}\tilde{P}_{\mathfrak{c}^\prime}}.
\end{equation}
Furthermore, in Eqs.~(\ref{eq:1CCE},\ref{eq:2CCE},\ref{eq:3CCE},\ref{eq:MCCE}%
), all of the denominators are equal to unity because the system is
initially prepared at $\vert0\rangle$, which is also the eigenstate of $H_{%
\mathrm{NV}}$.

Generally speaking, it is impossible to exactly calculate $P(t)$ for a large
number of spins as the dimension of Hilbert space scales exponentially with
the number of spins. The $M$-CCE method approximates $P(t)$ as
\begin{equation}
P^{(M)}=\prod_{\vert\mathfrak{c}\vert\leq M}\tilde{P}_{\mathfrak{c}},
\end{equation}
where $\vert\mathfrak{c}\vert$ is the number of spins in the cluster $%
\mathfrak{c}$. For example, the first-order truncation of $P(t)$ yields
\begin{equation}
P^{(1)}=\prod_i\tilde{P}_{\{i\}}.
\end{equation}
And the second-order truncation of $P(t)$ reads
\begin{equation}
P^{(2)}=\prod_i\tilde{P}_{\{i\}}\prod_{i,j}\tilde{P}_{\{i,j\}}.
\end{equation}

The magnetic dipole-dipole coupling between nearest neighbors are typically
of several kHz \cite{Zhao11-1}, which plays a role in a time scale of orders
longer than the QZE under consideration. Further numerical simulation, which
is not shown here, confirms that the diference between the quantum dynamics
of electron spin with and without magnetic dipole-dipole couplings is
indistinguishable. Therefore, we neglect the magnetic dipole-dipole
couplings between nuclear spins in order to simplify the simulation.

\section*{Acknowledgements}

We thank L.-P. Yang for helpful discussions. FGD was supported by the
National Natural Science Foundation of China under Grant No.~11474026 and
the Fundamental Research Funds for the Central Universities under Grant
No.~2015KJJCA01. WY was supported by the
National Natural Science Foundation of China under Grant No.~11274036 and No.~11322542, and the MOST under Grant No.~2014CB848700.
QA was supported by the National Natural Science Foundation of China under Grant No.~11505007, the Youth Scholars Program of Beijing Normal University under Grant No.~2014NT28, and the Open Research Fund Program of the State Key Laboratory of Low-Dimensional Quantum Physics, Tsinghua University under Grant No.~KF201502. MZ was supported by the National Natural Science Foundation of China under Grant No.~11475021.

\section*{Author contributions statement}

All authors wrote the main manuscript text. Z.S.Y. did the calculations. W.
Y. develops the generalized CCE approach to deal with NV electron spin
relaxation. Q.A. and F.G.D. designed the project. Q.A. supervised the whole
project. All authors reviewed the manuscript.

\section*{Additional information}

Competing financial interests: The authors declare no competing financial
interests.

\end{document}